\documentclass[preprint, authoryear]{elsarticle}

\usepackage{natbib}
\usepackage[utf8]{inputenc} 
\usepackage[T1]{fontenc}
\usepackage{lmodern}
\usepackage[fleqn]{mathtools}
\usepackage{rotating}
\usepackage{amssymb}
\usepackage{amsmath}
\usepackage{graphicx}
\usepackage{caption}
\usepackage{epstopdf}
\usepackage{stackrel}
\usepackage{soul}
\usepackage{color}
\usepackage{xcolor}
\usepackage{relsize,exscale}

\newcommand*\diff{\mathop{}\!\mathrm{d}}

\renewcommand*{\underline}{\ul}

\newdefinition{rmk}{Remark}
\newproof{pf}{Proof}
\newproof{pot}{Proof of Theorem \ref{thm2}}

\begin{document}

\title{Tail models and the statistical limit of accuracy\\ in risk assessment \\[12pt]}

\author[hhu]{Ingo Hoffmann\corref{cor1}}
\ead{Ingo.Hoffmann@hhu.de}
\author[hhu]{Christoph J.~B\"orner}
\ead{Christoph.Boerner@hhu.de}

\cortext[cor1]{Corresponding author. Tel.: +49 211 81-15258; Fax.: +49 211 81-15316}

\address[hhu]{Financial Services, Faculty of Business Administration and Economics, \\ Heinrich Heine University D\"usseldorf, 40225 D\"usseldorf,
	Germany}

\begin{abstract}
	In risk management, tail risks are of crucial importance.
	The assessment of risks should be carried out in accordance with the regulatory authority's requirement at high quantiles. In general, the underlying distribution function is unknown, the database is sparse, and therefore special tail models are used.
	Very often, the generalized Pareto distribution is employed as a basic model, and its parameters are determined with data from the tail area. With the determined tail model, statisticians then calculate the required high quantiles. In this context, we consider the possible accuracy of the calculation of the quantiles and determine the finite sample distribution function of the quantile estimator, depending on the confidence level and the parameters of the tail model, and then calculate the finite sample bias and the finite sample variance of the quantile estimator. Finally, we present an impact analysis on the quantiles of an unknown distribution function.
\end{abstract}

\begin{keyword}
	Exceedances \sep Extreme Value Theory \sep Generalized Pareto Distribution \sep Quantile Estimation \sep Risk Assessment \sep Tail Models. \\[10pt]
\end{keyword}

\maketitle

\thispagestyle{empty} 
\newpage
\section{Introduction}
In many disciplines, there is often a need to adapt a statistical model to existing data to be able to make statements regarding uncertain future outcomes. In particular, when assessing risks, an estimate of major losses must be based on events that, despite having a low probability of occurrence, have a high impact. Since the actual distribution of data -- the parent distribution -- is generally unknown, statisticians begin their modeling with a guess regarding the underlying statistical model. In a first step, they try to fit one or more parametric distribution functions as a model to the data to evaluate the rare events in the next step. These models generally do not perfectly reflect the data. However, specific statistical tests can be applied to assess how well or how poorly a model fits the data as a whole. Nevertheless, especially in the case of rare events and high damage, small uncertainties in the assumption of a model lead to a faulty description of these extremes and
will call into question the value of the information obtained by this approach. Therefore, any uncertainties regarding the underlying model and the resulting misjudgments must be assumed to negatively affect the quality of the models for rare events. In particular, model uncertainties pose problems in the finance and insurance industries, especially when rare events need to be evaluated, for example, by calculating high quantiles.\\

To make more precise statements regarding rare events and their severity, statisticians can describe the tail of the parent distribution function using a separate model and calculate the corresponding quantiles more precisely to improve the quality of the results. 
In the case of financial institutions, the respective regulatory frameworks provide statisticians and risk managers with the confidence levels of the parent distribution quantiles  \citep{basel04, directive09, directive13, regulation13}. Depending on the purpose, the confidence level is frequently given as 99.9\%; however, the available data often do not cover this area at all. 
The calculation of the capital that is regulatorily required to take a risk is based on the value-at-risk (VaR) or conditional value-at-risk (CVaR), which are calculated from high quantiles. 
In addition to these regulatory risk measures, additional measures exist for internal management decisions, such as the return on risk-adjusted capital (RORAC) and the risk-adjusted return on capital (RAROC), which are also calculated from high quantiles of the parent distribution and are important for the risk assessment of a company. Given this framework, standalone modeling of the tails of the underlying distribution is suggested for a more accurate calculation of the risk values.\\

For a very large class of parent distribution functions, the generalized Pareto distribution (GPD) can be used as a model for the tail, cf.,\ e.g.,\ \cite{embrechts03} and the large number of references on this topic contained therein. This class of distributions includes all common parent distributions that play a role in the financial sector such that almost no uncertainty exists regarding the model selection for the tail of the unknown parent distribution. The required quantiles can then be determined to high confidence levels with a sufficient certainty.\\

In practice, when modeling risks, further constraints need to be considered (e.g.,\ a small database). Moreover, 
the question arises: which accuracy can be achieved at all in the risk assessment?
Specifically, if in the ideal case only the statistical errors remain, what accuracy can be expected within the risk assessment? This question, which is central to financial companies and audit authorities, will be addressed in this study.\\

The question of the accuracy of a tail model is discussed in detail in the literature regarding the parameter estimators of the GPD \citep{smith84,choulakian01,embrechts03}. The focus is often on the statistical properties of the estimators and less on the statistical properties of the quantiles. 
\citet{hosking87} have compared different methods of parameter estimation for the GPD. Thereby, on the basis of Monte Carlo simulations and asymptotic functions for the variance of the quantile estimator, they could already make statements about the properties of the quantile estimator of the GPD.
In the field of climatology, it is known that quantile estimators are biased, but it seems that more detailed causal research has not been carried out. The research in the field of climatology focuses primarily on practical methods to reduce the bias \citep{hoffmannP18, fang15}. In this study, we go beyond the state of the art and aim to examine the statistical properties of the quantiles and consider the reason for the bias of the quantiles.	 \\

After a brief description of risk modeling and quantile estimation (Sec.\ \ref{RiskAssesment}), we derive a finite sample distribution of the quantile estimator of the GPD (Sec.\ \ref{finitdistribution}) and further explore the distribution in the context of limited datasets (Sec.\ \ref{biasvariance}). 
As a result, we find that the quantile estimator is positively biased for a finite number of data and grows larger with smaller amounts of data following a power law. The same behavior is found for the variance of the quantile estimator. The results also indicate and quantify that these inaccuracies increase for risks whose underlying unknown distribution function has a fat tail. Of particular interest could be the positive bias: quantiles and thus risks are systematically overestimated for small databases. For practical applications, we provide a correction formula to mitigate the impact of this overestimation (Sec.\ \ref{BiasCorrection}).
Overall, these inaccuracies are expected to affect all derived risk measures, based on the quantile estimator. 
The results of this study therefore illustrate the accuracy limit that can be achieved during risk assessment in practice, especially if the underlying parent distribution is unknown, the database is small, and the regulatory framework requires risk assessment with high quantiles.\\

Note that in the following, peculiarities of censorized or winzorized data or contaminations of these due to superimposed distribution functions are not taken into account.
Instead, in our further considerations, we exclude censoring and contamination of the data and assume that the dataset consists of independently and identically distributed realizations. 
Therefore, in the following, we focus clearly on financial data with these properties. In this way, we consider the best of all imaginable cases and further examine the achievable accuracy for finite data series.
\section{Risk assessment at high quantiles}\label{RiskAssesment}
Especially when high quantiles are considered in risk assessment in practice, a separate modeling of the tail of the generally unknown parent distribution often takes place. Predominantly, the GPD is used as a tail model in practice \citep{basel09}. 
In the following, the main features of the GPD are briefly considered, as well as their use in tail modeling. The estimation of high quantiles of an unknown underlying distribution function is attributed to the estimation of high quantiles with the GPD as a model. Furthermore, the databases occurring in practice are described as a framework condition.
\subsection{Model of the tail of a distribution}\label{GPD}
A theorem in extreme value theory, which goes back to \citet{gnedenko43}, \citet{balkema74} and \citet{pickands75}, states that for a broad class of distributions, the distribution of the excesses over a threshold $u$ converges to a GPD 
if the threshold is sufficiently large.\\

The GPD is usually expressed as a two-parameter distribution and has the following distribution function:
\begin{flalign}\label{GPDDF}
F(x) & = 1-\left(1+\xi\frac{x}{\sigma} \right)^{-\frac{1}{\xi}},
\end{flalign}
where $\sigma$ is a positive scale parameter and $\xi$ is a shape parameter (sometimes called the tail parameter). The density function is
\begin{flalign}\label{GPDPF}
f(x) & = \frac{1}{\sigma}\left(1+\xi\frac{x}{\sigma} \right)^{-\frac{1+\xi}{\xi}},
\end{flalign}
with support $0\leq x< \infty$ for $\xi\geq 0$ and $0\leq x\leq -\frac{\sigma}{\xi}$ when $\xi< 0$. 
The quantile function of the GPD depending on the confidence level $\alpha$ is:
\begin{flalign}\label{qfunction}
q_\alpha & = \frac{\sigma}{\xi}\Big[\big(1-\alpha \big)^{-\xi}-1\Big].
\end{flalign}
The mean and variance are ${\text E}[x] = \frac{\sigma}{1-\xi}$ and ${\text{Var}}[x] = \frac{\sigma^2}{(1-\xi)^2(1-2\xi)} $, respectively; thus, the mean and variance of the GPD are positive and finite only for $\xi < 1$ and $\xi < 0.5$, respectively. For special values of $\xi$, the GPD leads to various other distributions. When $\xi = 0$ and $-1$, the GPD becomes an exponential and a uniform distribution, respectively. For $\xi>0$, the GPD has a long tail to the right and is a reparameterized version of the usual Pareto distribution. Several areas of applied statistics have used the latter range of $\xi$ to model datasets that exhibit this form of a long tail. \\

Since the GPD was introduced by \citet{pickands75}, numerous theoretical advancements and applications have followed \citep{davison84, smith84, smith85, vanmontfort85, hosking87, davison90, choulakian01}. The applications of the GPD include use in the analysis of extreme events in hydrology, as a failure-time distribution in reliability studies and in the modeling of large insurance claims. Numerous examples of applications can be found in \citet{embrechts03} and the studies listed therein. The GPD is also increasingly used in the financial and banking sectors. Especially in the assessment of risks based on high quantiles, the GPD is one of the proposed distributions for modeling the tail of an unknown parent distribution \citep{basel09}.\\

The preferred method in the literature for estimating the parameters of the GPD is the well-studied maximum likelihood method \citep{davison84,smith84,smith85,hosking87}. \citet{choulakian01} stated that it is theoretically possible to have datasets for which no solution to the likelihood equations exists, and they concluded that, in practice, this is extremely rare. In many practical applications, the estimated shape parameter $\hat\xi$ ranges between -0.5 and 0.5, and a solution to the likelihood equations exists \citep{hosking87,choulakian01}. For practical and theoretical reasons, these authors limited their attention to this range of values. \citet{hoffmann18a} adapted the GPD as a model for the tail of different parent distributions applicable to finance and banking and also found that the maximum likelihood estimator $\hat\xi$ falls within this range. 
Furthermore, in many applications with real data from the finance sector, it was found that $\hat{\xi}$ was only slightly smaller than zero and was limited upward by $\hat{\xi}<0.5$ \citep{hoffmann18a}. Thus, the range of interest of the tail index in practice can be given by the interval $0.0 \lesssim \xi <0.5$, indicating that the (unknown) underlying parent distribution functions may have longer upper tails. 
\subsection{Determination of the model parameters}\label{ModelParameter}
Let $X_1, X_2, \ldots , X_N$ be a sample of random variables with common unknown continuous distribution function $F(x)$ and density function $f(x)$.
Let further $x_{(1)} \geq x_{(2)}\geq \ldots \geq x_{(N)}$ be the sample values (in ascending order) obtained by ordering each realization $x_1, x_2, \ldots , x_N$ of the random variables $X_1, X_2, \ldots , X_N$.\\

A specific value $x_{n+1}$ for $n=1,\ldots, N-1$ determined from the ordered sample then separates the data into two parts. The first part of the data belongs to the body of the unknown parent distribution function, and the second part belongs to the tail. The value $x_{n+1}$ is referred to as the threshold $u$ at which the tail begins. The threshold can be estimated, for example, by the method of \citet{hoffmann18a}: $\hat{u} = x_{\hat{n}+1}$, where $\hat{n} +1$ is an estimate of the index of the data point that marks the threshold.\\

The parameters $\xi$ and $\sigma$ of the GPD, Eq.\ (\ref{GPDDF}), as the tail model are then determined via maximum likelihood estimation for the ordered subset of the data: $x_{(1)} \geq x_{(2)}\geq \ldots \geq x_{(\hat{n})}$. This leads to the finite sample estimates $\hat{\xi}_{\hat{n}}$ and $\hat{\sigma}_{\hat{n}}$. To simplify the notation, we omit the index $\hat{n}$ below.
\subsection{Estimation of high quantiles.}\label{Highquantiles}
The estimation function for the quantile $Q_\alpha$ of the unknown parent distribution with a confidence level $\alpha$ can be noted as follows \citep{embrechts03}:
\begin{flalign}\label{QEstimator}
\hat{Q}_\alpha 
& = \hat{u} + \frac{\hat{\sigma}}{\hat{\xi}}\left[ \left( \frac{N}{\hat{n}}  (1-\alpha)\right)^{-\hat{\xi}} - 1\right]
\end{flalign}
where $\hat{\xi}$ and $\hat{\sigma}$ are the maximum likelihood estimates of the parameters of the GPD. Furthermore, $\hat{u}$ again denotes the estimator of the threshold after a data point $x_{\hat{n}}$ with estimated index $\hat{n}$. 
A quantile estimator $\hat{q}_\alpha $ is defined by substituting estimators $\hat{\xi}$ and $\hat{\sigma}$ for the parameters in Eq.\ (\ref{qfunction}), cf., e.g., \citet{hosking87}:
\begin{flalign}\label{qestimatorM}
\hat{q}_\alpha 
&= \frac{\hat{\sigma}}{\hat{\xi}}\Big[\big(1-\alpha\big)^{-\hat{\xi}} - 1\Big].
\end{flalign}
Then, we have
\begin{flalign}\label{alpha}
(1-\alpha)^{-\hat{\xi}}
&= 1+ \frac{\hat{\xi}}{\hat{\sigma}}\;\hat{q}_\alpha.
\end{flalign}
With this equation, we rewrite Eq.\ (\ref{QEstimator}), and it follows that a relation between the estimated quantile $\hat{q}_\alpha$ of the GPD and the estimated quantile $\hat{Q}_\alpha$ of the unknown parent distribution is:
\begin{flalign}\label{Qofq}
\hat{Q}_\alpha 
& = \hat{u} + \frac{\hat{\sigma}}{\hat{\xi}}\left[ 
\left(\frac{\hat{n}}{N} \right)^{\hat{\xi}}  -1 \right]
+ \left(\frac{\hat{n}}{N} \right)^{\hat{\xi}} \hat{q}_\alpha.
\end{flalign}
This show that the accuracy of the quantile estimator $\hat{Q}_\alpha$ is influenced by many factors. The estimator $\hat{n}$ and thus $\hat{u}$ can be set very well in a sample with the method described for example by \citet{hoffmann18a}. The estimators $\hat{\xi}$ and $\hat{\sigma}$ are maximum likelihood estimators with the properties of being consistent and asymptotically efficient \citep{embrechts03}, so that they converge in the limit of large samples against the true values. The question is: which statistical properties does $\hat{q}_\alpha$ have? This will be examined below. Knowing the statistical properties, it is possible to clarify how the inaccuracies in the quantile $\hat{q}_\alpha$ affect the quantile $\hat{Q}_\alpha$.
\subsection{Practical considerations -- restrictions due to the database.}
\label{restrictions}
The regulatory requirements stipulate that risks in the financial sector should be calculated at high confidence levels and thus high quantiles, whereby a certain holding period is assumed (weekly, fortnightly, monthly, quarterly or annually), cf.,\ e.g.,\ \citet{basel04, basel09, directive09, directive13}. The quantiles are determined on the basis of datasets for each risk category (e.g.,\ assets) individually or on the basis of a bundled risk (e.g.,\ a portfolio). To estimate the amount of usable data, let us consider an example: if the risks of assets are valued within the period since the introduction of the euro, less than $N \approx 5000$ data points per asset are available for the analysis based on daily closing prices. The latter is the ideal case. For an asset launched later, the database will be reduced accordingly. If the GPD is used to determine the high quantiles, the estimation of the parameters of the tail model is based on a much smaller part of the database.
Previous reports have already carried out studies on the favorable choice of the amount of data used for tail modeling in the finance and insurance fields \citep{mcneil97, moscadelli04, dutta07}, revealing that the preferred tail length $n$ for the data series analyzed in those fields comprises approximately 10\% to 15\% of the total amount of data available. Hence, the database for estimating $q_\alpha$ is limited in the majority of cases upwards by $n\approx 750$. If further restrictions are added -- for example, if there are only weekly data available -- the database is reduced again. 
On the other hand, for $n< 50$ data points, the influence of statistical errors increases enormously, so that for smaller data bases an evaluation of high quantiles becomes numerically difficult; see also Sec.\ \ref{biasvariance}.\\

In summary, in most cases, the number of data points used in tail modeling practice is somewhere in the interval between $n = 50 \ldots 1000$ data points. This amount of data is another restriction in tail modeling. This limitation follows from practice and will be considered in the further analysis.
\section{Density of the finite sample distribution of the quantile $q_\alpha$}
\label{finitdistribution}
Based on the results of \citet{smith87} and \cite{embrechts03}, we derive by a straightforward calculation a very good approximation of the density for the finite sample distribution of the quantile estimator $\hat{q}_\alpha$. The technical details of the calculation can be found in \ref{DensityDerivation}; here, we only note the final result:
\begin{flalign}\label{DensityOfq}\nonumber
& f_{q}(z; n, \alpha, \sigma, \xi) = \\ \nonumber
& \; \frac{1}{2\pi}\;\frac{n}{\sigma}\;\frac{1}{\sqrt{1+4\xi + 5\xi^2 + 2\xi^3}}\;\; \mathop{\mathlarger{\mathlarger{\mathlarger\int}}}_{-\infty}^{+\infty}\diff u \; \psi(u) \;\times\\
& \exp\left\{-\frac{n}{1+2\xi}\left[ \frac{\big(u-\xi\big)^2}{1+\xi} 
+ \frac{\big(u-\xi \big)\big(z\psi(u)-\sigma\big)}
{(1+\xi)\sigma}
+ \frac{\big(z\psi(u)-\sigma\big)^2}{2\sigma^2} 
    \right]\right\}
\end{flalign}
with $\psi(u) = \frac{u}{(1-\alpha)^{-u} -1}$ and
$n$ being the sample size; $\alpha$ is the confidence level, and $\sigma$ and $\xi$ are the actual parameters of the GPD.\\

\begin{figure}[htbp]
	\centering
	\captionsetup{labelfont = bf}
	\includegraphics[width=0.975\textwidth]{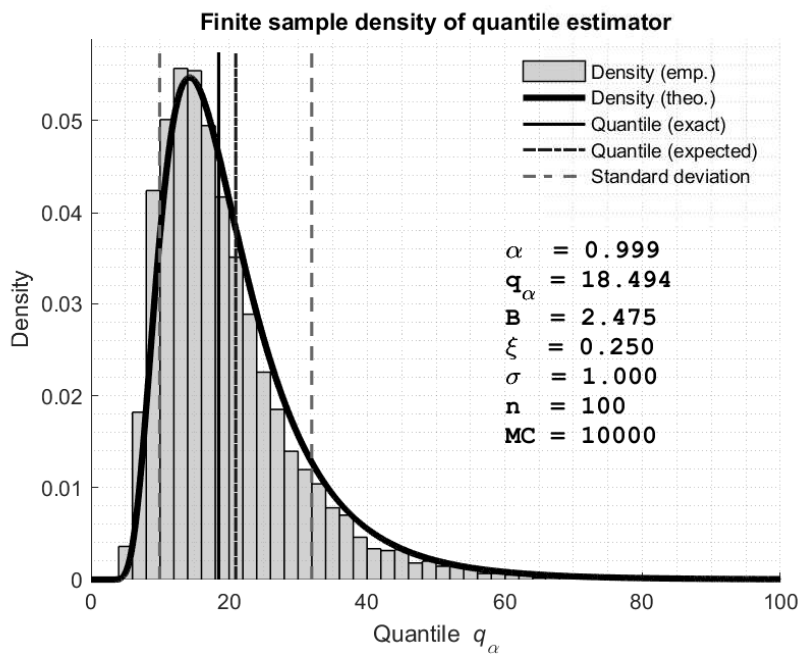}
	\begin{quote}
		\caption[Histogram]{\label{histogram} As an example, for an arbitrary parameter configuration, the finite sample density of the distribution of the quantile estimator $\hat{q}_\alpha$ is shown in comparison to the histogram of the empirically determined quantiles. In addition, the location of the statistical quantities $\text{E}[\hat{q}_\alpha]$ and $\sqrt{\text{Var}[\hat{q}_\alpha]}$ compared to the true quantile ${q}_\alpha = 18.494\ldots$ is shown.}
	\end{quote}
\end{figure}

As an example with an arbitrary parameter configuration, Fig.\ \ref{histogram} shows a comparison of the theoretical density, Eq.\ (\ref{DensityOfq}), with the corresponding empirical density.
In addition, the graph shows the location of the actual quantile $q_{0.999}$ and the expected value $\textrm{E}[\hat{q}_{0.999}]$ and the standard deviation of the quantile estimator when considering a sample of $n = 100$ data points. Note that the expected value is above the actual value. This observation means that the quantile estimator $\hat{q}_\alpha$ is positively biased, with $B = 2.475$.\\

For various parameter configurations, the performed goodness-of-fit tests do not reject the hypothesis that the theoretical density, Eq.\ (\ref{DensityOfq}), describes the empirical density. However, for very small sample lengths $n<50$ and as $\xi$ becomes significantly less than zero, this observation changes. 
The reason for this is that with this parameter configuration, the assumption of the asymptotic normal distribution \citep{smith87}, used by us, for the maximum likelihood estimators $\hat{\xi}$ and $\hat{\sigma}$ is clearly not justified; see also \ref{DensityDerivation}. 
However, this assumption is used to calculate the density $f_q$ and works well for a sample length of $n> 50$ and the parameter range $ \xi = 0.0, \ldots, 0.5$. 
The latter range covers all applications occurring in the financial sector where the measured data belong to a distribution function with a possibly fat tail. In addition, with this parameter interval, the mappable value range of the data is not limited by the tail model upwards; see also Sec.\ \ref{GPD}. Therefore, in the further investigations, we concentrate on this parameter range. We also assume that data scaling is possible in practice, so in the following sections we focus on the parameter setting $\sigma = 1$. Furthermore, we focus on the area of high quantiles, which is important for regulators and auditors, at a confidence level of $\alpha = 0.999$. However, the following calculation can also be performed with a different set of parameters.
\section{Finite sample bias and variance of the quantile estimator}\label{biasvariance}
In general, the bias $B$ is calculated as the expected value:
\begin{flalign}\label{BiasFormel}
B(n,\alpha, \sigma, \xi) = \textrm{E}[\hat{q}_\alpha - q_\alpha].
\end{flalign}
The actual quantile $q_\alpha$ and the estimator of the quantile $\hat{q}_\alpha$ depend on the confidence level $\alpha$, the scale parameter $\sigma$ and the shape parameter $\xi$.
Furthermore, the estimator of the quantile also depends on the sample length $n$.\\

\begin{figure}[htbp]
	\centering
	\captionsetup{labelfont = bf}
	\includegraphics[width=0.975\textwidth]{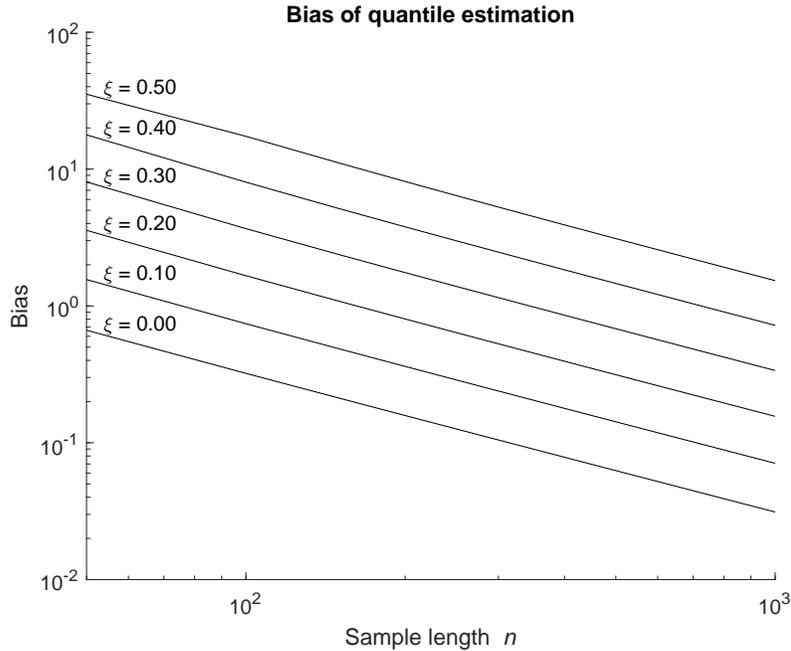}
	\begin{quote}
		\caption[Bias]{\label{Bias} The finite sample density, Eq.\ (\ref{DensityOfq}), of the quantile estimator is used to calculate the finite sample bias for datasets with $n = 50 \ldots 1000$ data points in length. For the underlying GPD, $\sigma = 1$ and 
		$\xi = 0.0, \ldots, 0.5$ are chosen.
		The parallel lines in the log-log plot indicate a power law with respect to the sample length $n$ and indicate exponential growth with respect to $\xi$ (see the text).}
	\end{quote}
\end{figure}

With $\sigma = 1$ and $\alpha = 0.999$, the expectation value in Eq.\ (\ref{BiasFormel}) is calculated with the density $f_q$ of Eq.\ (\ref{DensityOfq}) for a sample length of $n=50, \ldots, 1000$ and shape parameters $ \xi = 0.0, \ldots, 0.5$.
The finite sample bias $B(n,\xi) = B(n, 0.999, 1, \xi)$ calculated in this case is shown graphically in a log-log plot in Fig.\ \ref{Bias}.
As can be seen, the bias decreases with increasing sample length. This indicates that the quantile estimator converges to the actual quantile and has the statistical property of asymptotic consistency. We will examine this further in \ref{consistence}. There is also a clear dependence on the shape parameter. The larger the parameter $\xi$, the greater the bias is. This observation means that in practice, the fatter the tail, the greater the expected deviation of the quantile estimator is from the unknown actual quantile.\\

In the same way, with the density $f_q$, the variance of the quantile estimator can also be calculated. 
\begin{figure}[htbp]
	\centering
	\captionsetup{labelfont = bf}
	\includegraphics[width=0.975\textwidth]{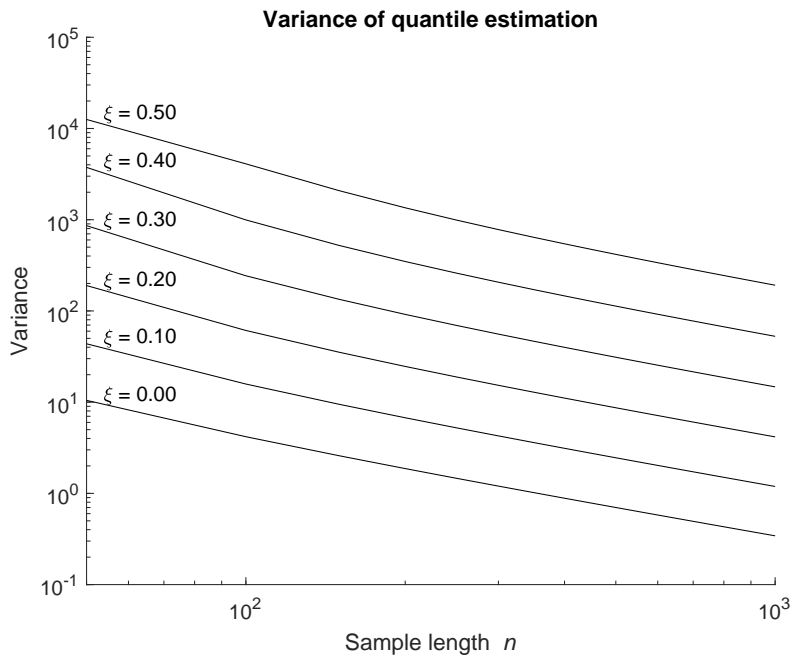}
	\begin{quote}
		\caption[Variance]{\label{Variance} The finite sample density, Eq.\ (\ref{DensityOfq}), of the quantile estimator is used to calculate the finite sample variance for data sets with $n = 50 \ldots 1000$ data points in length. For the underlying GPD, $\sigma = 1$ and 
		$\xi = 0.0, \ldots, 0.5$ are chosen. For smaller sample lengths, the power law changes to higher powers, and the variance as a function of the sample length in the log-log plot describes a convex curve. The parallel curves in the log-log plot indicate exponential growth with respect to $\xi$.}
	\end{quote}
\end{figure}
The log-log plot in Fig.\ \ref{Variance} shows the dependence of the variance on the modifiable parameters $n$ and $\xi$. It can be seen that the variance also converges to zero as the amount of data increases. In addition, as with the bias, it can be seen that there is a clear dependence on the shape parameter. The larger the parameter $\xi$, the greater the variance is. This observation means that in practice, the fatter the tail, the greater the variance of the quantile estimator is.\\

The results obtained here coincide with the observations of \citet{hosking87}. For certain parameter configurations, our results can be compared with their results; Table 4 (bias) and Table 5 (variance) in \citet{hosking87}. Although the methods are very different, numerically the results show a good agreement.\\

In practice, therefore, the estimate of the quantile should be subject to great uncertainties, especially for small datasets. In addition, the uncertainty increases when the whole dataset is based on an unknown distribution function that has a fat tail. The bias and the variance, which can be determined with the density $f_q$, are likely to influence every modeling approach and thus represent a limit of the modeling accuracy. 
If risk models are adjusted as best as possible, then this accuracy -- described by the calculated bias and the calculated variance -- can be achieved.
The calculated values -- bias and variance -- can be used as a benchmark for the evaluation of a specific modeling approach that aims to assess the risk at high quantiles.
\section{Correction of the bias of the quantile estimator}\label{BiasCorrection}
In Sec.\ \ref{biasvariance}, it was found that the quantile estimator $\hat{q}$ is systematically above the true value $q$ and thus has a certain positive bias $B$. 
To improve the risk assessment in practice, a method that corrects the bias of the quantile estimator would be useful. 
The conception of such a procedure is derived in this section from the previous results.
The correction method provides a way to mitigate in practice the influence of the bias.
\subsection{Quantile bias correction -- state of the art} 
\label{BiasCorrectionStatOfTheArt}
In the field of climatology, the necessity of bias corrections, especially corrections of the quantile estimator, has been discussed for a long time.
Therefore, there are already a great number of methods in climatology to correct the bias of different estimators such as the estimator of a certain quantile, cf.,\ e.g., \citet{schmidli06, sun11, themessl11, fang15, jeon16, hoffmannP18} and the related overwhelming references cited therein. Additionally, in the financial industry, with the tightening of regulatory requirements, such topics could become even more present. The question is whether the simple transfer of existing methods of climatology into the financial sector is fruitful or whether another approach is appropriate.\\

Recently, two methods \citep{hoffmannP18} have prevailed in climatology from the original five methods \citep{fang15} and are often used to correct the bias of the quantile.
\begin{enumerate}[A:]
	\item {\it Local intensity scaling} described in \citep{schmidli06, hoffmannP18}.
	The basic idea is an adjustment of the mean value of a simulated (or measured) finite sample and a reproduction of an ''adequate'' mean value, i.e.,\ a comparison value from reference data. 
	From the comparison of the two mean values, a correction factor is calculated, 
	and the individual measured or simulated values are corrected accordingly. Implicitly,
	the quantile is also corrected.
	
	\item {\it Analytical quantile mapping} described in \citep{sun11, themessl11, hoffmannP18}.	
	The basic idea is that for simulated (or measured) data and for reference data, the GPD is adjusted in the tail area as an analytical distribution function.
	From the deviation of the two tail models, an analytical transfer function is determined. With the transfer function, the individual measured or simulated values are corrected accordingly, and as before, implicitly, the quantile is also corrected.
\end{enumerate}

The methods described have been developed for specific purposes under certain conditions that are not necessarily found in the financial sector. The methods are therefore unlikely to be transferred to problems in the financial sector. 
For example, there are usually no reference data available in the financial area, and if so, other reference data or other reference periods will provide different corrections of the bias of the quantile estimator. 
In addition, the ''manipulation'' of the measured data should also be viewed very critically by the statistician and the regulatory authorities, because other values determined from the dataset also change or the connection between different statistical quantities is disturbed.
The latter problems are similarly discussed in climatology \citep{hoffmannP18}. However, an essential point is not taken into account in the described methods and has not yet been considered to the best of the authors' knowledge: the finite number of data points itself already causes a bias in the quantile estimator. This bias can be determined analytically (see Sec.\ \ref{biasvariance}) and should be taken into account in further risk assessments, especially in finance. This will be considered below. 
From Eq.\ (\ref{Qofq}), it can be seen that the bias of $\hat{q}_\alpha$ has an effect on $\hat{Q}_\alpha$.
If the bias is determined for $\hat{q}_\alpha$, an impact analysis on the estimator $\hat{Q}_\alpha$ can be carried out in practice in individual cases.
\subsection{Quantile bias correction --  a formula for practice}
\label{BiasCorrectionFormula}
In Fig.\ \ref{Bias}, the bias is shown as a function of the sample size $n$ and the pre-given shape parameter $\xi$. The figure shows only a small number of parallel graphs. In fact, $\xi$ was varied in smaller steps between 0.0 and 0.5.
With the calculated data points $B$ as a function of the sample size $n$ and the shape parameter $\xi$, we performed nonlinear regression. The formula for the basic model can be read from the graphic and has the following form:
\begin{flalign}\label{basicmodel}
B(n, \xi) & = n^{a_1} \exp\big\{(\ln 10) \left(  a_2\xi +a_3  \right) \big\}.
\end{flalign}
Nonlinear regression yielded the following parameters:
\begin{flalign*}
a_1  = -1.00733 \qquad a_2  = +3.49572 \qquad a_3  = +1.49397
\end{flalign*}
For reasons of simplification, the parameters are rounded. Then, the following formula for the bias of the quantile estimator can be used for a practical application:
\begin{flalign}\label{parcticbias}
B(n, \xi) & = \frac{1}{n}\exp\left\{ \frac{\ln 10}{2} \Bigg( 7\xi +3  \Bigg) \right\}.
\end{flalign}
Hence, an estimated quantile $\hat{q}_{0.999}$ can be shifted by $\tilde{q}_{0.999} = \hat{q}_{0.999} - B(n, \hat{\xi})$. This should in practice lead to a new estimator $\tilde{q}_{0.999}$ of the quantile closer to the actual, generally unknown value of the quantile ${q}_{0.999}$.\\

It should be noted that Eq.\ (\ref{parcticbias}) applies to a confidence level of $\alpha = 0.999$ and a scale parameter of $\sigma = 1$. Other configurations can be similarly calculated using the finite sample density of the quantile estimator Eq.\ (\ref{DensityOfq}). With $B(n,\alpha, \sigma, \xi) = \textrm{E}[\hat{q}_\alpha] - q_\alpha $ and $\tilde{q}_\alpha = \hat{q}_\alpha - B(n, \alpha, \sigma, \hat{\xi})$, it follows that $\textrm{E}[\tilde{q}_\alpha] = q_\alpha$.\\

Finally, an important property can be observed from Eq.\ (\ref{parcticbias}): As $n \rightarrow\infty$, the bias becomes 0 -- independent of $\xi$ -- indicating again that the quantile estimator $\hat{q}_\alpha$ is asymptotically consistent and converges to the true value of the quantile ${q}_\alpha$. 
In fact, it can be shown theoretically in a simple way that the estimator $\hat{q}_\alpha$ of the quantile of the GPD is asymptotically consistent. Technical details of the calculation are shown in \ref{consistence}.
\section{Discussion and Conclusion}\label{Discussion}
In financial practice, there is a regulatory need to estimate quantiles at high confidence levels from measured data ({\it\textsf{iid}}) with an unknown distribution function. In this case, a GPD is usually adapted as a tail model for the unknown distribution function for a subset of the measured data. The required quantile $\hat{Q}_\alpha$ is then calculated with the inverse function of the estimated GPD, Eq.\ (\ref{QEstimator}), or likewise with the estimated quantile $\hat{q}_\alpha$, Eq.\ (\ref{Qofq}).\\

In this article, we considered the statistical property of the quantile estimator $\hat{q}_\alpha$. 
The investigations focused on the parameter range $\xi = 0.0,\ldots, 0.5$, which is important for risk assessment. The GPD then has the support $\mathbb{R}^{\geq 0}$ and can be used as a model for unknown distribution functions that have a fat tail.\\

The starting point of our analyses was the density of the finite sample distribution of the quantile estimator for the quantiles of the GPD. Thus, the finite sample variance and the finite sample bias of the quantile estimator could be determined. Further, we showed that the quantile estimator is asymptotically consistent and converges to the true value for large datasets. For practical applications, an approximate correction formula, Eq.\ (\ref{parcticbias}), was derived, which can mitigate the negative influence of the bias on the estimated quantile of an unknown distribution function.\\

Generally, the results show that in practice, for finite, small datasets, the quantile $\hat{q}_\alpha$ determined is positive biased ($B > 0$) and has considerable uncertainty ($\textrm{Var} [\hat{q}_\alpha ] \gg 0$ as $n<50$). 
The calculation shown is universal, so that for other confidence levels or scale parameters the bias and the variance of the quantile estimator can be calculated.
This may allow new perspectives for the review of risk assessment procedures and should also be of interest to audit authorities and regulators.\\

The results presented represent a lower limit of the accuracy in quantile estimation and can be used in other works as an absolute benchmark for the quality of the quantile estimate, e.g.,\ in automated procedures for threshold detection and tail modeling, cf.,\ e.g.,\ \citet{hoffmann18a} and the references cited therein.
Future theoretical work will address the complete statistical properties of the quantile estimator $\hat{Q}_\alpha$, Eq.\ (\ref{Qofq}). In another branch, the examination can be extended to the parameter area where the GPD has compact support, i.e., $\xi = -0.5, \ldots, 0.0$.
\appendix
\section{Derivation of the density $f_{q}(z; n, \alpha, \sigma, \xi)$}
\label{DensityDerivation}
The goal is to obtain an approximation of the finite sample density for the estimator of the quantile $\hat{q}_\alpha$ of the GPD, Eq.\ (\ref{GPDDF}). As a starting point, we take the results of \citet{smith87} and \citet{embrechts03} in the notation of the latter author.
Let $\hat{\sigma}$ and $\hat{\xi}$ be the finite sample maximum likelihood estimators for the actual scale parameter $\sigma$ and the actual shape parameter $\xi > -0.5$. Both
estimators are determined for a sample of length $n$.
Then, the distribution of the random vector ${\bf X}=\sqrt{n} (\hat{\xi} - \xi, \frac{\hat{\sigma}}{\sigma} - 1)'$ converges to a centered normal distribution with covariance matrix:
\begin{flalign}\label{Mmatrix}
{\bf M}^{-1} & = (1 + \xi)
\begin{pmatrix}
1 + \xi & -1 \\ -1  &  2
\end{pmatrix}.
\end{flalign}
This means that ${\bf X} \stackrel{d}{\rightarrow} {\cal N} ({\bf 0}, {\bf M}^{-1})$ if $n \rightarrow \infty$.\\

Defining vector ${\bf a} = -\sqrt{n}(\xi, 1)'$ and matrix ${\bf B} = \sqrt{n}\,\textrm{diag}(1 , \frac{1}{\sigma})$, the vector ${\bf Y} = (\hat{\xi}, \hat{\sigma})'$ is given by the affine map: ${\bf Y} = {\bf B}^{-1}({\bf X} - {\bf a})$. Then, the distribution of the random vector ${\bf Y}$ converges to a normal distribution with parameters:
\begin{flalign} \label{Normalparameter}
\boldsymbol{\mu} & = (\xi, \sigma)' \quad\textrm{ and } \quad {\bf C}  = \frac{(1 + \xi)}{n}
\begin{pmatrix}
1 + \xi & -\sigma \\ -\sigma  &  2\sigma^2
\end{pmatrix}.
\end{flalign}
Thus, ${\bf Y} \stackrel{d}{\rightarrow} {\cal N} (\boldsymbol{\mu}, {\bf C})$ if $n \rightarrow \infty$. Hence, the probability density function of the vector ${\bf Y}$ for finite $n$ is approximately given by the density $f_{\bf Y}(u, v)$ of a bivariate normal distribution with the parameters given in Eq.\ \ref{Normalparameter}. Note that $u$ and $v$ represent the integration variables used later for $\hat{\xi}$ and $\hat{\sigma}$.\\

Standard methods, cf.,\ e.g., \citet{kendall76}, then lead us by a straightforward calculation to the distribution function $\Phi(\hat{q}_\alpha)$ of the estimator of the quantile. By performing a double integration of $f_{\bf Y}(u, v)$ within a region $\cal D$, the sought density for the quantile estimator can be obtained from the resulting formula.\\

The integration domain $\cal D$ is determined from the estimator of the quantile:
\begin{flalign}\label{qestimator}
\hat{q}_\alpha 
= \frac{\hat{\sigma}}{\hat{\xi}}\Big[\big(1-\alpha\big)^{-\hat{\xi}} - 1\Big].
\end{flalign}
Thus, the double integration is carried out as follows:
\begin{flalign}\label{CDFofq1}
\Phi(\hat{q}_\alpha) 
& = \mathop{\int}_{-\infty}^{+\infty}\diff u 
\mathop{\int}_{-\infty}^{\hat{q}_\alpha \psi(u)} \diff v \; f_{\bf Y}(u, v),
\end{flalign}
with $\psi(u) = \frac{u}{(1-\alpha)^{-u}-1}$. After a suitable substitution -- introducing a new integration variable $z$ -- and an exchange of the order of integration, 
\begin{flalign}\label{CDFofq2}
\Phi(\hat{q}_\alpha) 
& = \mathop{\int}_{-\infty}^{\hat{q}_\alpha } \diff z
\mathop{\int}_{-\infty}^{+\infty}\diff u \; \psi(u)  f_{\bf Y}(u, z \psi(u))  = \mathop{\int}_{-\infty}^{\hat{q}_\alpha } \diff z \; f_q(z).
\end{flalign}
Simple algebraic transformations lead to the representation of the desired density $f_q(z)$ shown in Eq.\ (\ref{DensityOfq}).

\section{Asymptotic consistence of the quantile estimator}\label{consistence}
The above results in \ref{DensityDerivation} make it easy to show that the estimator $\hat{q}_\alpha$ of the quantile of the GPD is asymptotically consistent.\\

First, we notice that with the renormalized matrix ${\tilde{\bf C}} = n {\bf C} $ of the covariance matrix, Eq.\ (\ref{Normalparameter}), and its Cholesky decomposition ${\tilde{\bf C}}^{-1} = \big({\tilde{\bf C}}^{-\frac{1}{2}}\big)'\big({\tilde{\bf C}}^{-\frac{1}{2}}\big)$, the density $f_{\bf Y}(u, v)$ of the bivariate normal distribution becomes a bivariate Dirac delta function for $n\rightarrow\infty$. Using the algebraic properties of the delta function, cf.,\ e.g.,\ \cite{oldham09}, one quickly obtains:
\begin{flalign}\label{bivariatdeltadistribution}
f_{\bf Y}(u, v) = \delta(u - \xi) \delta(v - \sigma) \quad\textrm{ for }\quad n\rightarrow\infty,
\end{flalign}
where $\xi$ and $\sigma$ are the actual parameters of the GPD. 
Substituting Eq.\ (\ref{bivariatdeltadistribution}) into Eq.\ (\ref{CDFofq2}) leads to:
\begin{flalign}
f_q(z) & = \mathop{\int}_{-\infty}^{+\infty}\diff u \; \psi(u) \delta(u - \xi) \delta(z\psi(u) - \sigma) = \delta(z - q_\alpha).
\end{flalign}
This means that the distributions of the estimate $\hat{q}_\alpha$ of the quantile become more and more concentrated near the true value of the quantile $q_\alpha$ being estimated, if $n\rightarrow\infty$, so that the probability of the estimator being arbitrarily close to $q_\alpha$ converges to one.

\end{document}